\documentclass[prb,twocolumn,showpacs]{revtex4}
\usepackage{amsmath}
\usepackage{dcolumn}
\usepackage{graphicx}

\begin{document}

\title[Short title for running header]{Seeing the orbital ordering in Iron-based superconductors with magnetic anisotropy}
\author{Yuehua Su$^{1}$ and Tao Li$^{2}$}
\affiliation{$^{1}$Department of Physics, Yantai University,
Yantai 264005, P.R.China\\ $^{2}$Department of Physics, Renmin
University of China, Beijing 100872, P.R.China}
\date{\today}

\begin{abstract}
The orbital fluctuation of the conduction electrons in the
Iron-based superconductors is found to contribute significantly to
the magnetic response of the system. With the use of a realistic
five-band model and group theoretical analysis, we have determined
the orbital magnetic susceptibility in such a multi-orbital system.
At $n=6.1$, the in-plane orbital magnetic susceptibility is
predicted to be about 10$\mu_{\mathrm{B}}^{2}/\mathrm{eV}$, which is
more than $2/3$ of the observed total susceptibility around 200 K in
122 systems(of about 14$\mu_{\mathrm{B}}^{2}/\mathrm{eV}$ or
$4.5\times10^{-4}\mathrm{erg}/\mathrm{G}^{2}\mathrm{mol}_{\mathrm{AS}}$\cite{Klingeler}).
We find the in-plane orbital magnetic response is sensitive to the
breaking of the tetragonal symmetry in the orbital space. In
particular, when the observed band splitting(between the $3d_{xz}$
and the $3d_{yz}$-dominated band) is used to estimate the strength
of the symmetry breaking perturbation\cite{Shen}, a 4.5\% modulation
in the in-plane orbital magnetic susceptibility can be produced,
making the latter a useful probe of the orbital ordering in such a
multi-orbital system. As a by product, the theory also explains the
large anisotropy between the in-plane and the out-of-plane magnetic
response observed universally in susceptibility and NMR
measurements.
\end{abstract}


\maketitle

An unresolved issue in the study of the Iron-based superconductors
is the role of their multi-orbital nature. In most other
superconductors, the orbital degree of freedom is quenched at low
energy in the crystal field environment. However, both LDA
calculation and ARPES measurement\cite{Feng,Shen,Shimojima} indicate
that in the Iron-based superconductors all the five Fe $3d$ orbital
play essential role in forming the low energy degree of freedom
around the Fermi surface. Many novel properties of the Iron-based
superconductors, especially those in the name of electronic
nematicity\cite{Chuang,ChuPRB,Chu1,Chu2,Kasahara}, have been argued
to be related to the orbital ordering in these
systems\cite{Ku,Lv1,Singh,Lv2,Nevidomskyy}. Most recently, a
two-fold modulation of the magnetic susceptibility in the Fe-Fe
plane is found to develop around a temperature that is significantly
higher than the structural phase transition point\cite{Kasahara}.
However, it is still a mystery how the observed electronic
nematicity is related to the orbital ordering of the system.

Another puzzle about the Iron-based superconductors is the strong
anisotropy in their magnetic response observed universally in
susceptibility and Knight shift
measurements\cite{ChenCa,ChenBa,CanfieldSr,Zheng}. The
susceptibility in the Fe-Fe plane is found to be significantly
larger than that perpendicular to it. This is very unusual, since
the magnetic response of a transition metal is usually attributed to
the spin of its valence electron and is essentially isotropic. The
orbital magnetic response, on the other hand, is usually quenched as
a result of the crystal field effect. However, since the crystal
field splitting in the Iron-based superconductors is very small and
all the five $3d$ orbital are involved in the low energy
physics\cite{Feng,Shen,Shimojima}, the orbital angular momentum of
the conduction electron can contribute to the magnetic response of
these systems. Such a contribution is intrinsically anisotropic and
depends on the electronic structure of the system, especially on the
symmetry breaking in the orbital space.

The purpose of this paper is to evaluate orbital magnetic response
of the Iron-based superconductors from a realistic model and to
explore the relation between orbital ordering and the electronic
nematicity observed in recent torque magnetometry
measurement\cite{Kasahara}. We find the orbital magnetic
susceptibility in these multi-orbital systems is comparable in
magnitude with the measured total magnetic susceptibility. More
specifically, the in-plane orbital magnetic susceptibility is
predicted to be about 10$\mu_{\mathrm{B}}^{2}/\mathrm{eV}$, which
accounts for more than $2/3$ of the observed susceptibility at 200 K
in 122 systems\cite{Klingeler,ChenCa,ChenBa,CanfieldSr}.
Furthermore, the in-plane orbital magnetic response is found to be
sensitive to the breaking of the tetragonal symmetry in the orbital
space, making it a useful probe of orbital ordering in these
multi-orbital systems. As a by product, the observed strong
anisotropy between the in-plane and out-of-plane magnetic
susceptibility also find a natural explanation from our calculation.

\begin{figure}[h!]
\includegraphics[width=9cm,angle=0]{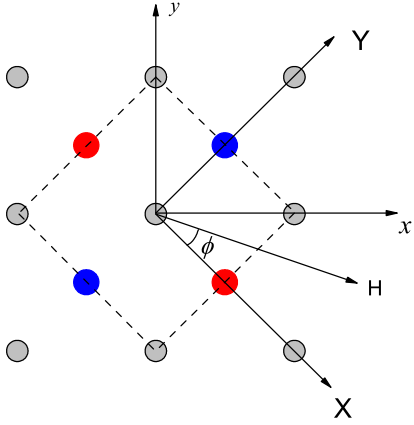}
\caption{The square lattice of the Fe ions (shown as gray dots) and
the local coordinate system for the atomic orbital. The red and blue
dots denote the As ions above and below the Fe-Fe plane. $\phi$ is
the angle between the $X$-axis and the direction in which magnetic
susceptibility is measured. In the tetragonal phase, the point group
symmetry around the Fe ion is $D_{2d}$, which is broken down to
$D_{2}$ in the orthogonal phase. } \label{fig1}
\end{figure}

The Iron-based superconductors have a very complicated band
structure. In this study, we adopt the five-band tight-binding model
derived from fitting the LDA band structure\cite{Kuroki} of the
LaFeAsO system. Following the notations of Ref.\onlinecite{Kuroki},
the band model reads,
\begin{eqnarray}
H_{kin}=\sum_{i,j}\sum_{\nu,\nu',\sigma}[t_{i,j}^{\nu,\nu'}c_{i,\nu,\sigma}^{\dagger}c_{j,\nu',\sigma}+h.c.]
 +\sum_{i,\nu,\sigma}\varepsilon_{\nu}n_{i,\nu,\sigma}, \nonumber \\
 \label{eqn1}
\end{eqnarray}
where $\nu,\nu'=1,..,5$ is the index for the five maximally
localized Wannier functions(MLWFs) on the Fe site, namely,
$|1\rangle=|3d_{3Z^{2}-R^{2}}\rangle$, $|2\rangle=|3d_{XZ}\rangle$,
$|3\rangle=|3d_{YZ}\rangle$, $|4\rangle=|3d_{X^{2}-Y^{2}}\rangle$
and $|5\rangle=|3d_{XY}\rangle$. $t_{i,j}^{\nu,\nu'}$ denotes the
hopping integral between the $\nu$-th and $\nu'$-th orbital at site
$i$ and site $j$. Here an unfolded scheme is adopted as in
Ref.\onlinecite{Kuroki}. The $X$ and $Y$-axis for the Wannier
functions, which are in the Fe-As bond direction, are rotated by 45
degree from the $x$ and $y$-axis of the Fe-Fe square lattice(see
Fig.\ref{fig1}). $\varepsilon_{\nu}$ is the on-site energy of the
$\nu$-th orbital. The hopping integral is truncated at the fifth
neighbor and the values of the model parameters can be found in
Ref.\onlinecite{Kuroki}.

The interaction of electron has the following general form
\begin{eqnarray}
 &H_{int}&=U\sum_{i,\nu,\sigma}n_{i,\nu,\sigma}n_{i,\nu,\overline{\sigma}}
 +(U'-J)\sum_{i,\nu' \neq \nu,\sigma}n_{i,\nu,\sigma}n_{i,\nu',\sigma}\nonumber\\
 &+U'&\sum_{i,\nu' \neq
 \nu,\sigma}n_{i,\nu,\sigma}n_{i,\nu',\overline{\sigma}}-J\sum_{i,\nu'\neq
 \nu,\sigma}c^{\dagger}_{i,\nu,\sigma}c_{i,\nu,\overline{\sigma}}c^{\dagger}_{i,\nu',\overline{\sigma}}c_{i,\nu',\sigma}\nonumber\\
 &-J&\sum_{i,\nu' \neq
 \nu,\sigma}c^{\dagger}_{i,\nu,\sigma}c^{\dagger}_{i,\nu,\overline{\sigma}}c_{i,\nu',\sigma}c_{i,\nu',\overline{\sigma}}.\label{eqn2}
\end{eqnarray}
Here we have included the intra- and inter-orbital Coulomb
repulsion, the Hund's rule coupling and the pair hopping term and
have assumed that $U'=U-2J$.
$n_{i}=\sum_{\nu,\sigma}n_{i,\nu,\sigma}=\sum_{\nu,\sigma}c^{\dagger}_{i,\nu,\sigma}c_{i,\nu,\sigma}$
is the number density operator of the electron and
$\overline{\sigma}=-\sigma$.

Since the five Fe $3d$ orbital $|\nu\rangle$ are all real functions,
they can not carry current and thus their orbital angular momentum
are quenched in the static limit. However, since the orbital content
varies on the Fermi surface, fluctuation in the orbital character
and orbital angular momentum survives in the low energy limit and
can contribute to the magnetic response of the system. In the
following we will calculate such a magnetic response in the RPA
scheme.

The orbital magnetic susceptibility is defined through the
correlation function of the orbital magnetic moment in the following
way
\begin{eqnarray}
\chi^{\alpha}_{L}({\mathbf q},\tau)=-\langle
T_{\tau}\hat{L}^{\alpha}({\mathbf q},\tau)\hat{L}^{\alpha}(-{\mathbf
q},0) \rangle ,\label{eqn4}
\end{eqnarray}
in which $\hat{L}^{\alpha}({\mathbf q},\tau)$ denotes the Fourier
component of the orbital magnetic moment density in the $\alpha$
direction and $\alpha=X,Y,Z$. Here we use $\mu_{B}^{2}$ as the unit
of susceptibility. The operator for the orbital magnetic moment on a
given site is defined as
$\hat{L}^{\alpha}=\sum_{\nu,\nu',\sigma}c^{\dagger}_{\nu,\sigma}l^{\alpha}_{\nu,\nu'}c_{\nu',\sigma}$,
where $l^{\alpha}_{\nu,\nu'}$ is the matrix element of the orbital
magnetic moment in the basis spanned by the five MLWFs.

The matrix element $l^{\alpha}_{\nu,\nu'}$ can be determined in
principle from a first principle calculation. Here we will be
satisfied with the result of a semi-quantitative analysis, for which
much simplification can be achieved when symmetry arguments are
adopted. In the following, we will illustrate the steps for
$l^{Z}_{\nu,\nu'}$. First, since $\hat{L}^{Z}$ is time reversal odd
and the five $3d$ orbital are all real, $l^{Z}_{\nu,\nu'}$ must be
purely imaginary. Second, since $\hat{L}^{Z}$ is odd under the
action of the three generators of the $D_{2d}$ point group around
each Fe ion, namely $R_{x}(\pi)$, $\sigma_{X}$ and
$\sigma_{Y}$\cite{group}, while the five $3d$ orbital transform as
\begin{eqnarray}
R_{x}(\pi): \left(\begin{array}{c}
        |1\rangle \\
        |2 \rangle \\
        |3 \rangle \\
        |4 \rangle \\
        |5\rangle \\
      \end{array}\right)\rightarrow
\left(\begin{array}{c}
        |1\rangle \\
        -|3 \rangle \\
        -|2 \rangle \\
        -|4 \rangle \\
        |5\rangle \\
      \end{array}\right)\nonumber
\end{eqnarray}
\begin{eqnarray}
\sigma_{X}: \left(\begin{array}{c}
        |1\rangle \\
        |2 \rangle \\
        |3 \rangle \\
        |4 \rangle \\
        |5\rangle \\
      \end{array}\right)\rightarrow
\left(\begin{array}{c}
        |1\rangle \\
        -|2 \rangle \\
        |3 \rangle \\
        |4 \rangle \\
        -|5\rangle \\
      \end{array}\right)\nonumber
\end{eqnarray}
and
\begin{eqnarray}
\sigma_{Y}:\left(\begin{array}{c}
        |1\rangle \\
        |2 \rangle \\
        |3 \rangle \\
        |4 \rangle \\
        |5\rangle \\
      \end{array}\right)\rightarrow
\left(\begin{array}{c}
        |1\rangle \\
        |2 \rangle \\
        -|3 \rangle \\
        |4 \rangle \\
        -|5\rangle \\
      \end{array}\right),\nonumber
\end{eqnarray}
the only none-zero matrix elements are $l^{Z}_{2,3}=-l^{Z}_{3,2}$
and $l^{Z}_{4,5}=-l^{Z}_{5,4}$. Thus $\hat{L}^{Z}$ can be generally
written as
\begin{equation}
\hat{L}^{Z}=i\sum_{\sigma}[\gamma_{1}
c^{\dagger}_{2,\sigma}c_{3,\sigma}+\gamma_{2}
c^{\dagger}_{4,\sigma}c_{5,\sigma}]+h.c.,\nonumber
\end{equation}
in which $\gamma_{1}$ and $\gamma_{2}$ are two real numbers.
Following the same line of reasoning one find that
\begin{eqnarray}
\hat{L}^{X}&=i&\sum_{\sigma}[\gamma_{3} c^{\dagger}_{1,\sigma}c_{3,\sigma}+\gamma_{4} c^{\dagger}_{2,\sigma}c_{5,\sigma}+\gamma_{5} c^{\dagger}_{3,\sigma}c_{4,\sigma}]+h.c.\nonumber\\
\hat{L}^{Y}&=-i&\sum_{\sigma}[\gamma_{3}
c^{\dagger}_{1,\sigma}c_{2,\sigma}+\gamma_{4}
c^{\dagger}_{3,\sigma}c_{5,\sigma}-\gamma_{5}
c^{\dagger}_{2,\sigma}c_{4,\sigma}]+h.c.\nonumber,
\end{eqnarray}
with the three real coefficients $\gamma_{3,4,5}$ left undetermined.

To have an estimate of the values of the five coefficients
$\gamma_{1,..,5}$, we approximate the five MLWFs $|\nu\rangle$,
$\nu=1,..,5$, with the five Fe $3d$ orbital in the atomic limit.
These atomic orbital are related to the spherical harmonics of $l=2$
in the following ways(apart from the radial part of the wave
function which is not used in determining the matrix element of
$\hat{L}^{\alpha}$)
\begin{eqnarray}
|1\rangle&=&|2,0\rangle\nonumber\\
|2\rangle&=&\frac{1}{\sqrt{2}}\left(|2,-1\rangle-|2,1\rangle\right)\nonumber\\
|3\rangle&=&\frac{i}{\sqrt{2}}\left(|2,-1\rangle+|2,1\rangle\right)\nonumber\\
|4\rangle&=&\frac{1}{\sqrt{2}}\left(|2,-2\rangle+|2,2\rangle\right)\nonumber\\
|5\rangle&=&\frac{i}{\sqrt{2}}\left(|2,-2\rangle-|2,2\rangle\right),
\nonumber
\end{eqnarray}
where $|2,m\rangle\propto Y^{m}_{2}$ are the spherical harmonics of
$l=2$. Since $\langle2,m'|\hat{L}^{Z}|2,m\rangle=m\delta_{m,m'}$ and
$\langle2,m|\hat{L}^{+}|2,m'\rangle=\sqrt{6-m'(m'+1)}\delta_{m,m'+1}$(here
$\hat{L}^{+}=\hat{L}^{X}+i\hat{L}^{Y}$), we have
\begin{eqnarray}
\gamma_{1}&=-&\gamma_{4}=\gamma_{5}=-1\nonumber\\
\gamma_{2}&=&-2\nonumber\\
\gamma_{3}&=&\sqrt{3}\nonumber.
\end{eqnarray}
We will use these values in the following calculation.

The bare orbital magnetic susceptibility is readily obtained as
\begin{eqnarray}
\chi^{0,\alpha}_{L}(T)&=&\lim_{{\mathbf q}\rightarrow
0}\frac{2}{N}{\sum_{\mathbf{k},m,m'}}\frac{f(\xi_{\mathbf{
k+q},m'})-f(\xi_{{\mathbf k},m})} {\xi_{{\mathbf k},m}-\xi_{\mathbf{
k+q},m'}} \left|L^{\alpha}_{{\mathbf
k},m,m'}\right|^{2}.\nonumber\\
\label{eqn6}
\end{eqnarray}
Here $\xi_{{\mathbf k},m}=\epsilon_{{\mathbf k},m}-\mu$ is the band
energy of the $m$-th band($m=1,...,5$) and $\mu$ is the chemical
potential. $L^{\alpha}_{{\mathbf
k},m,m'}=\sum_{\nu,\nu'}l^{\alpha}_{\nu,\nu'}u^{*}_{{\mathbf
k},\nu,m}u_{{\mathbf k},\nu',m'}$ and $u_{{\mathbf k},\nu,m}$ is the
$m$-th eigenvector of the band Hamiltonian at momentum ${\mathbf
k}$. As a comparison, the Pauli spin susceptibility is given by
\begin{eqnarray}
\chi^{0,\alpha}_{S}(T)&=&\lim_{{\mathbf q}\rightarrow
0}\frac{2}{N}\sum_{{\mathbf k},m}\left[\frac{f(\xi_{\mathbf{
k+q},m})-f(\xi_{{\mathbf k},m})} {\xi_{{\mathbf k},m}-\xi_{\mathbf{
k+q},m}}\right]\nonumber.
\end{eqnarray}
Unlike the orbital magnetic susceptibility, the Pauli spin
susceptibility has contribution only from intra-band process. Thus
at low temperature the spin susceptibility is solely determined by
the electronic state around the Fermi surface, while the orbital
magnetic susceptibility depends on electronic states both on and far
away from the Fermi energy. As a result, both the temperature and
the doping dependence of the orbital magnetic response should be
much weaker than that of the spin magnetic response.

Now we consider the RPA correction of the orbital magnetic
susceptibility. The orbital magnetic excitation of the system has
the general form of
$\hat{O}^{\nu\nu'}=i\sum_{\sigma}(c^{\dagger}_{\nu,\sigma}c_{\nu',\sigma}-c^{\dagger}_{\nu',\sigma}c_{\nu,\sigma})$.
Without losing generality, we assume $\nu'>\nu$. There are in total
10 such excitations and all of them are time reversal odd and spin
singlet. The correlation function between these excitations can be
defined in the following way
\begin{equation}
\chi^{\nu\nu',\upsilon\upsilon'}_{O}({\mathbf q},\tau)=-\langle
T_{\tau}\hat{O}^{\nu\nu'}({\mathbf
q},\tau)\hat{O}^{\upsilon\upsilon'}(-{\mathbf q},0) \rangle
,\nonumber
\end{equation}
and the corresponding bare susceptibility in the static limit
$\chi^{0,\nu\nu',\upsilon\upsilon'}_{O}(T)$ is given by an
expression similar to Eq.(\ref{eqn6}), except that the matrix
element $ |L^{\alpha}_{{\mathbf k},m,m'}|^{2}$ should be replaced by
\begin{eqnarray}
O^{\nu\nu',\upsilon\upsilon'}_{{\mathbf k},m,m'}& =&(u^{*}_{{\mathbf
k},\nu',m'}u_{{\mathbf k},\nu,m}-u^{*}_{{\mathbf
k},\nu,m'}u_{{\mathbf k},\nu',m})\nonumber\\
&\times&(u^{*}_{{\mathbf k},\upsilon,m}u_{{\mathbf
k},\upsilon',m'}-u^{*}_{{\mathbf k},\upsilon',m}u_{{\mathbf
k},\upsilon,m'}).\nonumber
\end{eqnarray}

The RPA correction of $\chi^{\nu\nu',\upsilon\upsilon'}_{O}$ is
contributed by the inter-orbital Coulomb repulsion, the Hund's rule
coupling and the pair hopping term. The RPA kernel is extremely
simple and is given by
$V_{\nu\nu',\upsilon\upsilon'}=\frac{(U'-J)}{4}\delta_{\nu\nu',\upsilon\upsilon'}$(see
Supplementary material A). The RPA corrected susceptibility can be
written formally as
\begin{equation}
\chi_{O}=\frac{\chi^{0}_{O}}{1-V\chi^{0}_{O}},\nonumber
\end{equation}
in which $\chi_{O}$, $\chi^{0}_{O}$ and $V$ are all to be understood
as $10\times10$ matrix(we note while $V$ is a diagonal matrix in the
space of $\hat{O}^{\nu,\nu'}$, $\chi^{0}_{O}$ is not). The orbital
magnetic susceptibility can be obtained from the combinations of the
matrix element of $\chi_{O}$. For example,
\begin{eqnarray}
\chi^{Z}_{L}=\chi^{23,23}_{O}
+4(\chi^{45,45}_{O}+\chi^{23,45}_{O}).\nonumber
\end{eqnarray}
The orbital magnetic susceptibility in other direction can be
obtained in a similar way.

The observation of the two-fold modulation in the in-plane magnetic
susceptibility indicates that the tetragonal symmetry of the system
is broken down to orthogonal. This can happen either through orbital
ordering, or through nematicity in spin
correlation\cite{Xu,Chubukov1}. Here we assume it happens through
orbital ordering, since the orbital magnetic response is much more
sensitive to it than to spin nematicity. The form of the symmetry
breaking perturbation in the orthogonal phase can be largely
determined by group theoretical arguments. Among the five $3d$
orbital, the $3d_{3Z^{2}-R^{2}}$, $3d_{XY}$ and $3d_{X^{2}-Y^{2}}$
orbital each form a one dimensional representation of the $D_{2d}$
point group. The $3d_{XZ}$ and $3d_{YZ}$ orbital form a
two-dimensional representation which becomes reducible when the
symmetry is lowered to orthogonal. We thus focus on symmetry
breaking terms in the space spanned by the $3d_{XZ}$ and $3d_{YZ}$
orbital. A group theoretical analysis then shows that up to nearest
neighboring hopping terms, the only allowable symmetry breaking
perturbation in the orthogonal phase takes the form (see
Supplementary information B)
\begin{eqnarray}
\Delta
H&=&\eta_{1}\sum_{i,\sigma}(c^{\dagger}_{i,2,\sigma}c_{i,3,\sigma}+c^{\dagger}_{i,3,\sigma}c_{i,2,\sigma})\nonumber\\
&+&\eta_{2}\sum_{i,\delta,\sigma}\mathrm{d}_{\delta}(c^{\dagger}_{i,2,\sigma}c_{i+\delta,2,\sigma}+c^{\dagger}_{i,3,\sigma}c_{i+\delta,3,\sigma})\nonumber\\
&+&\eta_{3}\sum_{i,\delta,\sigma}(c^{\dagger}_{i,2,\sigma}c_{i+\delta,3,\sigma}+c^{\dagger}_{i,3,\sigma}c_{i+\delta,2,\sigma}),
\label{eqn7}
\end{eqnarray}
in which $\delta=\pm x,\pm y$ is the vector between nearest
neighboring Fe sites. $\mathrm{d}_{\delta}$ is the d-wave form
factor and $\mathrm{d}_{\pm x}=1$, $\mathrm{d}_{\pm y}=-1$. Here,
$\eta_{1}$ is the strength of the on-site symmetry breaking
perturbation. $\eta_{2}$ and $\eta_{3}$ are the strengths of the
d-wave intra-orbital and s-wave inter-orbital hopping terms between
nearest neighboring Fe sites. From ARPES measurement\cite{Shen}, it
is found that the splitting between the $3d_{xz}$ and the
$3d_{yz}$-dominated band is zero at the $\Gamma$ point and maximizes
at the X and Y point. Among the three perturbations in
Eq.(\ref{eqn7}), only the d-wave intra-orbital hopping term is
consistent with such a momentum dependence. For example, both the
$\eta_{1}$ or $\eta_{3}$-type perturbation would result in an
nonzero band splitting at the $\Gamma$ point, which is not observed.
Furthermore, the $\eta_{3}$-type perturbation has no effect at the X
and Y point, where the observed band splitting reaches its maximum.
We thus set $\eta_{1}=\eta_{3}=0$. This leaves us $\eta_{2}$ as the
only undetermined parameter.

\begin{figure}[h!]
\includegraphics[width=8cm,angle=0]{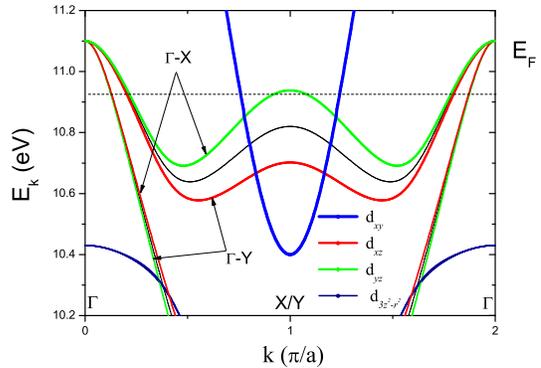}
\caption{Overlay of the band dispersion along the
$\Gamma-\mathrm{X}$ and $\Gamma-\mathrm{Y}$ direction in the
orthogonal phase. The orbital character is indicated by the color of
the lines and the dispersion in the tetragonal phase is plotted in
thin lines for reference. In the calculation we have set
$\eta_{2}=30$ meV. The dashed line indicates the Fermi level at
$n=6.1$. } \label{fig2}
\end{figure}

We are now at the position to present the numerical results. Our
calculation is done at a fixed band filling of $n=6.1$. The chemical
potential is determined by solving the mean field particle number
equation at each temperature. We have set $U=1.2$eV, $J=0.15$eV, as
is chosen in Ref.\onlinecite{Kuroki}. To estimate the value of
$\eta_{2}$ from the observed band splitting, we note that the band
width of the Iron-based superconductors is significantly smaller
than the prediction of band structure calculation. We thus fit the
relative rather than the absolute magnitude of the band splitting.
According to ARPES measurement, the maximal band splitting between
the $3d_{yz}$ and $3d_{xz}$-dominated band is about one half of the
dispersion of the $3d_{yz}$-dominated band between the $\Gamma$ and
X point\cite{Shen}. To fit such a splitting, we set $\eta_{2}=30$
meV. The calculated band dispersion along the $\Gamma-\mathrm{X}$
and $\Gamma-\mathrm{Y}$ direction is shown in Fig.\ref{fig2}, which
looks very similar to the experimental result\cite{Shen}. The
temperature dependence of $\eta_{2}$ is modeled by the mean field
form of $\eta_{2}(T)=\eta_{2}(0)\sqrt{1-(T/T_{c})^{2}}$, in which
$T_{c}$ is to be understood as the mean field critical temperature
of orbital ordering. We set $T_{c}= 150 $K in our
calculation\cite{Kasahara}.

In the tetragonal phase, the orbital magnetic susceptibility is
found to be isotropic in the Fe-Fe plane and is almost temperature
and doping independent for $6.0\leq n \leq 6.2$(see Supplementary
material C). This is reasonable since the orbital magnetic response
is contributed by the whole band, rather than the electronic state
near the Fermi level only. At $n=6.1$, the bare orbital magnetic
susceptibility in the Fe-Fe plane is found to be about
7.3$\mu_{\mathrm{B}}^{2}/\mathrm{eV}$, which is enhanced to
10$\mu_{\mathrm{B}}^{2}/\mathrm{eV}$ after RPA correction. This is
already comparable to the observed total in-plane magnetic
susceptibility at 200K in 122 systems, which is about
$4.5\times10^{-4}\mathrm{erg}/\mathrm{G}^{2}\mathrm{mol}_{\mathrm{AS}}$
(or 14$\mu_{\mathrm{B}}^{2}/\mathrm{eV}$)\cite{Klingeler}. As a
comparison, the bare Pauli spin susceptibility is only about
2$\mu_{\mathrm{B}}^{2}/\mathrm{eV}$.

\begin{figure}[h!]
\includegraphics[width=11cm,angle=0]{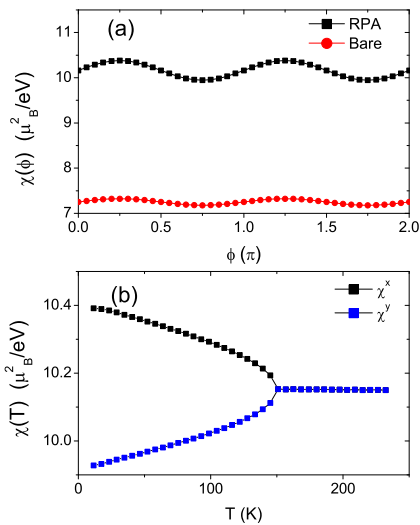}
\caption{(a)The in-plane modulation of the orbital magnetic
susceptibility before and after RPA correction. (b) The temperature
dependence of the RPA-corrected orbital magnetic susceptibility
along the two principle axes of the orthogonal phase.} \label{fig3}
\end{figure}

When a symmetry breaking perturbation of the $\eta_{2}$-type is
turned on, a two-fold modulation shows up in the in-plane orbital
magnetic susceptibility. The angular dependence of the in-plane
susceptibility at $T=T_{c}/5$ is shown in Fig.\ref{fig3}a. Here
$\phi$ denotes the angle between the $X$-axis and the direction in
which the magnetic susceptibility is measured. The relative strength
of the modulation is about 2.6\% before RPA correction and is
enhanced to 4.5\% after RPA correction. The principle axes of the
modulation are along the direction of the nearest Fe-Fe bond, which
is just what we should expect from our model construction. The
temperature dependence of the susceptibility in the principle axes
are shown in Fig.\ref{fig3}b. These predictions are in good
agreement with the result of the recent torque magnetometry
measurement\cite{Kasahara}. Thus the magnetic anisotropy provide a
realistic probe of the orbital ordering in the Iron-based
superconductors.

A robust prediction of our theory is the strong anisotropy between
the in-plane and the out-of-plane orbital magnetic susceptibility.
At $n=6.1$, the bare orbital magnetic susceptibility in $Z$
direction is found to be about
3.8$\mu_{\mathrm{B}}^{2}/\mathrm{eV}$, which is enhanced to
4.5$\mu_{\mathrm{B}}^{2}/eV$ after RPA correction. This is only
about the half of the value of the in-plane orbital magnetic
susceptibility. We find the ratio between the in-plane and
out-of-plane orbital magnetic susceptibility is also almost
temperature and doping independent for $6.0\leq n \leq 6.2$ and is
always close to 2. According to experiments, both the in-plane and
the out-of-plane magnetic susceptibility exhibit linear temperature
dependence with almost the same slope. However, the intercept of the
in-plane magnetic susceptibility is always much larger than that of
the out-of-plane magnetic
susceptibility\cite{ChenCa,ChenBa,CanfieldSr,Zheng}. This behavior
can be easily understood if we decompose the measured magnetic
susceptibility into an isotropic component that is linearly
temperature dependent and a temperature independent component that
is anisotropic, or,
\begin{equation}
\chi^{\alpha}(T)=\chi^{\alpha}_{L}+\chi_{S}(T).
\end{equation}
It is then quite natural to associate the anisotropic component
$\chi^{\alpha}_{L}$ with the orbital magnetic response, which is
essentially temperature independent. The isotropic component
$\chi_{S}(T)$ should then be attributed to the spin magnetic
response, whose linear temperature dependence is still an unresolved
issue in the field.\cite{GMZhang,SPKou,Chubukov2}

In our calculation, we have used a five-band model derived from the
band structure of the LaFeAsO system. However, the best known
susceptibility data on single crystalline sample are all taken from
the 122 system. It is thus better to perform the calculation with a
material-specific band structure for the 122 systems. While this is
an interesting possibility and should be pursued in the future, we
note that the basic structure of the bands in both the 1111 and the
122 systems are quite similar. Since the orbital magnetic response
is contributed by the whole band rather than the electronic state
near the Fermi level only, we expect the 1111 and 122 system to
exhibit similar orbital magnetic response. Another way to improve
our calculation is to use the matrix element of $\hat{L}^{\alpha}$
calculated from first principle code, rather than approximating them
with those in the basis spanned by the atomic orbital. However,
since the form the matrix element is largely determined by symmetry,
we do not expect such more advanced calculation to change the
conclusion of this paper in a qualitative way. Indeed, we find that
our results are not sensitive to the small variation of the
parameters $\gamma_{1,.,5}$.

In summary, we have shown that the orbital angular momentum of the
conduction electrons in the Iron-based superconductors contributes
significantly to the magnetic response of the system. In particular,
the theory predicts that the orbital magnetic susceptibility
accounts for more than $2/3$ of the observed magnetic susceptibility
at 200 K in 122 systems. We show that the orbital magnetic response
is sensitive to symmetry breaking in the orbital space, which makes
it a useful probe of the orbital ordering in these multi-orbital
systems.  A large and temperature independent anisotropy between the
in-plane and the out-of plane susceptibility is predicted, which
provides a natural understanding on the behavior of the magnetic
response of these systems.

Yuehua Su is support by NSFC Grant No. 10974167 and Tao Li is
supported by NSFC Grant No. 10774187, No. 11034012 and National
Basic Research Program of China No. 2010CB923004. We are grateful to
K. Kuroki for clarifying the phase convention used in
Ref.\onlinecite{Kuroki}.

\section{Supplementary materials}
\subsection{The form of the RPA kernel for orbital magnetic excitations}
The ten orbital magnetic excitation of the form
$\hat{O}^{\nu\nu'}=i\sum_{\sigma}(c^{\dagger}_{\nu,\sigma}c_{\nu',\sigma}-c^{\dagger}_{\nu',\sigma}c_{\nu,\sigma})$
are all time reversal odd and spin rotational invariant. In the
absence of time reversal symmetry breaking they form a subspace
within the space of all orbital excitations. It is thus sufficient
to restrict our consideration in this subspace.

The RPA correction to the orbital magnetic response is contributed
by the inter-orbital Coulomb term, the Hund's rule coupling term and
the pair hopping term. For example, the inter-orbital Coulomb term
has the following mean field decoupling($\nu'>\nu$),
\begin{eqnarray}
U'n_{i,\nu,\sigma}n_{\i,\nu',\sigma}\sim -&U' &\langle
c^{\dagger}_{i,\nu,\sigma}c_{i,\nu',\sigma}\rangle
c^{\dagger}_{i,\nu',\sigma}c_{i,\nu,\sigma}\nonumber\\
-&U' &\langle c^{\dagger}_{i,\nu',\sigma}c_{i,\nu,\sigma}\rangle
c^{\dagger}_{i,\nu,\sigma}c_{i,\nu',\sigma}\nonumber\\
+&U' &\langle c^{\dagger}_{i,\nu,\sigma}c_{i,\nu',\sigma}\rangle
\langle c^{\dagger}_{i,\nu',\sigma}c_{i,\nu,\sigma}\rangle.\nonumber
\end{eqnarray}
When expressed in terms of $\hat{O}^{\nu,\nu'}$, we have
\begin{eqnarray}
U'\sum_{\sigma}n_{i,\nu,\sigma}n_{\i,\nu',\sigma}\sim -\frac{U'}{4}
\langle \hat{O}^{\nu,\nu'}\rangle \hat{O}^{\nu,\nu'} +\frac{U'}{8}
\langle \hat{O}^{\nu,\nu'}\rangle \langle
\hat{O}^{\nu,\nu'}\rangle.\nonumber
\end{eqnarray}
Thus the RPA kernel is diagonal in the subspace of
$\hat{O}^{\nu,\nu'}$. Following the same steps, it can be shown that
the RPA correction contributed by the last two terms in
Eq.(\ref{eqn2}) cancels with each other.

\subsection{The form of the symmetry breaking perturbation in the orthogonal phase}
The form of the symmetry breaking perturbation in the orthogonal
phase can be determined from the following group theoretical
arguments. We first consider the form of the on-site symmetry
breaking term. The point group around each Fe ion in the orthogonal
phase is $D_{2}$ and has four one dimensional irreducible
representations. Among the five MLWFs, $|3Z^{2}-R^{2}\rangle$ and
$|XY\rangle$ both belong to the identity representation,
$|X^{2}-Y^{2}\rangle$ belongs to the $\mathrm{B}_{1}$
representation, the linear combinations $|XZ\rangle+|YZ\rangle$ and
$|XZ\rangle-|YZ\rangle$ belong to the $\mathrm{B}_{2}$ and
$\mathrm{B}_{3}$ representation. Thus symmetry allowed on-site
Fermion bilinear terms have the general form of
\begin{eqnarray}
H_{2}&=&\sum_{i,\sigma}(\beta_{1}c^{\dagger}_{i,1,\sigma}c_{i,1,\sigma}+\beta_{2}c^{\dagger}_{i,5,\sigma}c_{i,5,\sigma}+\beta_{3}c^{\dagger}_{i,4,\sigma}c_{i,4,\sigma})\nonumber\\
&+&\beta_{4}\sum_{i,\sigma}(c^{\dagger}_{i,1,\sigma}c_{i,5,\sigma}+c^{\dagger}_{i,5,\sigma}c_{i,1,\sigma})\nonumber\\
&+&\beta_{5}\sum_{i,\sigma}(c^{\dagger}_{i,2,\sigma}+c^{\dagger}_{i,3,\sigma})(c_{i,2,\sigma}+c_{i,3,\sigma})\nonumber\\
&+&\beta_{6}\sum_{i,\sigma}(c^{\dagger}_{i,2,\sigma}-c^{\dagger}_{i,3,\sigma})(c_{i,2,\sigma}-c_{i,3,\sigma})\label{eqn8}
\end{eqnarray}

In the tetragonal phase, the local symmetry around each Fe ion is
promoted to $D_{2d}$, which has four one dimensional representations
and a two dimensional representation. Among the five MLWFs,
$|3Z^{2}-R^{2}\rangle$ belongs to the identity representation,
$|XY\rangle$ and $|X^{2}-Y^{2}\rangle$ belong to the
$\mathrm{B}_{1}$ and $\mathrm{B}_{2}$ representation, the linear
combinations $|XZ\rangle+|YZ\rangle$ and $|XZ\rangle-|YZ\rangle$
form the two components of the two dimensional representation. For
this reason, the bilinear form
$c^{\dagger}_{i,1,\sigma}c_{i,1,\sigma}$,
$c^{\dagger}_{i,4,\sigma}c_{i,4,\sigma}$,
$c^{\dagger}_{i,5,\sigma}c_{i,5,\sigma}$, and
$c^{\dagger}_{i,2,\sigma}c_{i,2,\sigma}
+c^{\dagger}_{i,3,\sigma}c_{i,3,\sigma}$ all belong to the identity
representation of $D_{2d}$. When these symmetric perturbations are
removed from Eq.(\ref{eqn8}), we get the symmetric breaking
perturbation in the orthogonal phase, which now takes the form of
\begin{eqnarray}
\Delta
H&=&\lambda_{1}\sum_{i,\sigma}(c^{\dagger}_{i,2,\sigma}c_{i,3,\sigma}+c^{\dagger}_{i,3,\sigma}c_{i,2,\sigma})\nonumber\\
&+&\lambda_{2}\sum_{i,\sigma}(c^{\dagger}_{i,1,\sigma}c_{i,5,\sigma}+c^{\dagger}_{i,5,\sigma}c_{i,1,\sigma}),\nonumber
\end{eqnarray}
in which $\lambda_{1}=\beta_{5}-\beta_{6}$, $\lambda_{2}=\beta_{4}$.

\begin{figure}[h!]
\includegraphics[width=8cm,angle=0]{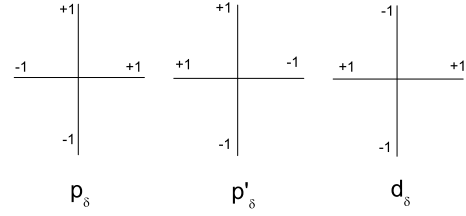}
\caption{An illustration of the p-wave and d-wave form factor
defined in the main text.} \label{sfig1}
\end{figure}

The above argument can be easily generalized to determined the form
the symmetry breaking perturbation on various bonds. In particular,
we find there are in total 13 independent symmetry breaking
perturbations on nearest neighboring Fe-Fe bonds. The form of these
terms are
\begin{eqnarray}
\Delta H=\Delta H_{s}+\Delta H_{p}+\Delta H_{d},\nonumber
\end{eqnarray}
in which
\begin{eqnarray}
\Delta H_{s}&=&\kappa_{2}\sum_{i,\delta,\sigma}(c^{\dagger}_{i,2,\sigma}c_{i+\delta,3,\sigma}+c^{\dagger}_{i,3,\sigma}c_{i+\delta,2,\sigma})\nonumber\\
&+&\sum_{i,\delta,\sigma}(\kappa_{3}c^{\dagger}_{i,1,\sigma}c_{i+\delta,5,\sigma}+\kappa_{4}c^{\dagger}_{i,5,\sigma}c_{i+\delta,1,\sigma})\nonumber
\end{eqnarray}
\begin{eqnarray}
\Delta H_{p}&=&\kappa_{8}\sum_{i,\delta,\sigma}(\mathrm{p}_{\delta}c^{\dagger}_{i,2,\sigma}c_{i+\delta,1,\sigma}+\mathrm{p}'_{\delta}c^{\dagger}_{i,3,\sigma}c_{i+\delta,1,\sigma})\nonumber\\
&+&\kappa_{9}\sum_{i,\delta,\sigma}(\mathrm{p}_{\delta}c^{\dagger}_{i,1,\sigma}c_{i+\delta,2,\sigma}+\mathrm{p}'_{\delta}c^{\dagger}_{i,1,\sigma}c_{i+\delta,3,\sigma})\nonumber\\
&+&\kappa_{10}\sum_{i,\delta,\sigma}(\mathrm{p}'_{\delta}c^{\dagger}_{i,2,\sigma}c_{i+\delta,5,\sigma}+\mathrm{p}_{\delta}c^{\dagger}_{i,3,\sigma}c_{i+\delta,5,\sigma})\nonumber\\
&+&\kappa_{11}\sum_{i,\delta,\sigma}(\mathrm{p}'_{\delta}c^{\dagger}_{i,5,\sigma}c_{i+\delta,2,\sigma}+\mathrm{p}_{\delta}c^{\dagger}_{i,5,\sigma}c_{i+\delta,3,\sigma})\nonumber\\
&+&\kappa_{12}\sum_{i,\delta,\sigma}(\mathrm{p}_{\delta}c^{\dagger}_{i,2,\sigma}c_{i+\delta,4,\sigma}-\mathrm{p}'_{\delta}c^{\dagger}_{i,3,\sigma}c_{i+\delta,4,\sigma})\nonumber\\
&+&\kappa_{13}\sum_{i,\delta,\sigma}(\mathrm{p}_{\delta}c^{\dagger}_{i,4,\sigma}c_{i+\delta,2,\sigma}-\mathrm{p}'_{\delta}c^{\dagger}_{i,4,\sigma}c_{i+\delta,3,\sigma}),\nonumber
\end{eqnarray}
and
\begin{eqnarray}
\Delta
H_{d}&=&\kappa_{1}\sum_{i,\delta,\sigma}\mathrm{d}_{\delta}(c^{\dagger}_{i,2,\sigma}c_{i+\delta,2,\sigma}+c^{\dagger}_{i,3,\sigma}c_{i+\delta,3,\sigma})\nonumber\\
&+&\kappa_{5}\sum_{i,\delta,\sigma}\mathrm{d}_{\delta}c^{\dagger}_{i,1,\sigma}c_{i+\delta,1,\sigma}\nonumber\\
&+&\kappa_{6}\sum_{i,\delta,\sigma}\mathrm{d}_{\delta}c^{\dagger}_{i,4,\sigma}c_{i+\delta,4,\sigma}\nonumber\\
&+&\kappa_{7}\sum_{i,\delta,\sigma}\mathrm{d}_{\delta}c^{\dagger}_{i,5,\sigma}c_{i+\delta,5,\sigma}.\nonumber
\end{eqnarray}
Here $\mathrm{p}_{\delta}$, $\mathrm{p'}_{\delta}$ are p-wave form
factors, $\mathrm{d}_{\delta}$ is the d-wave form factor. The value
of these form factors are illustrated in Fig.\ref{sfig1}

If we restrict our consideration to the subspace spanned by the
$d_{XZ}$ and $d_{YZ}$ orbital, then up to nearest neighboring
hopping term, the only allowable symmetry breaking perturbation has
the following form
\begin{eqnarray}
\Delta
H&=&\eta_{1}\sum_{i,\sigma}(c^{\dagger}_{i,2,\sigma}c_{i,3,\sigma}+c^{\dagger}_{i,3,\sigma}c_{i,2,\sigma})\nonumber\\
&+&\eta_{2}\sum_{i,\delta,\sigma}\mathrm{d}_{\delta}(c^{\dagger}_{i,2,\sigma}c_{i+\delta,2,\sigma}+c^{\dagger}_{i,3,\sigma}c_{i+\delta,3,\sigma})\nonumber\\
&+&\eta_{3}\sum_{i,\delta,\sigma}(c^{\dagger}_{i,2,\sigma}c_{i+\delta,3,\sigma}+c^{\dagger}_{i,3,\sigma}c_{i+\delta,2,\sigma}),\nonumber
\end{eqnarray}
in which $\eta_{1}=\lambda_{1}=\beta_{5}-\beta_{6}$,
$\eta_{2}=\kappa_{1}$, $\eta_{3}=\kappa_{2}$.

\begin{figure}[h!]
\includegraphics[width=7cm,angle=0]{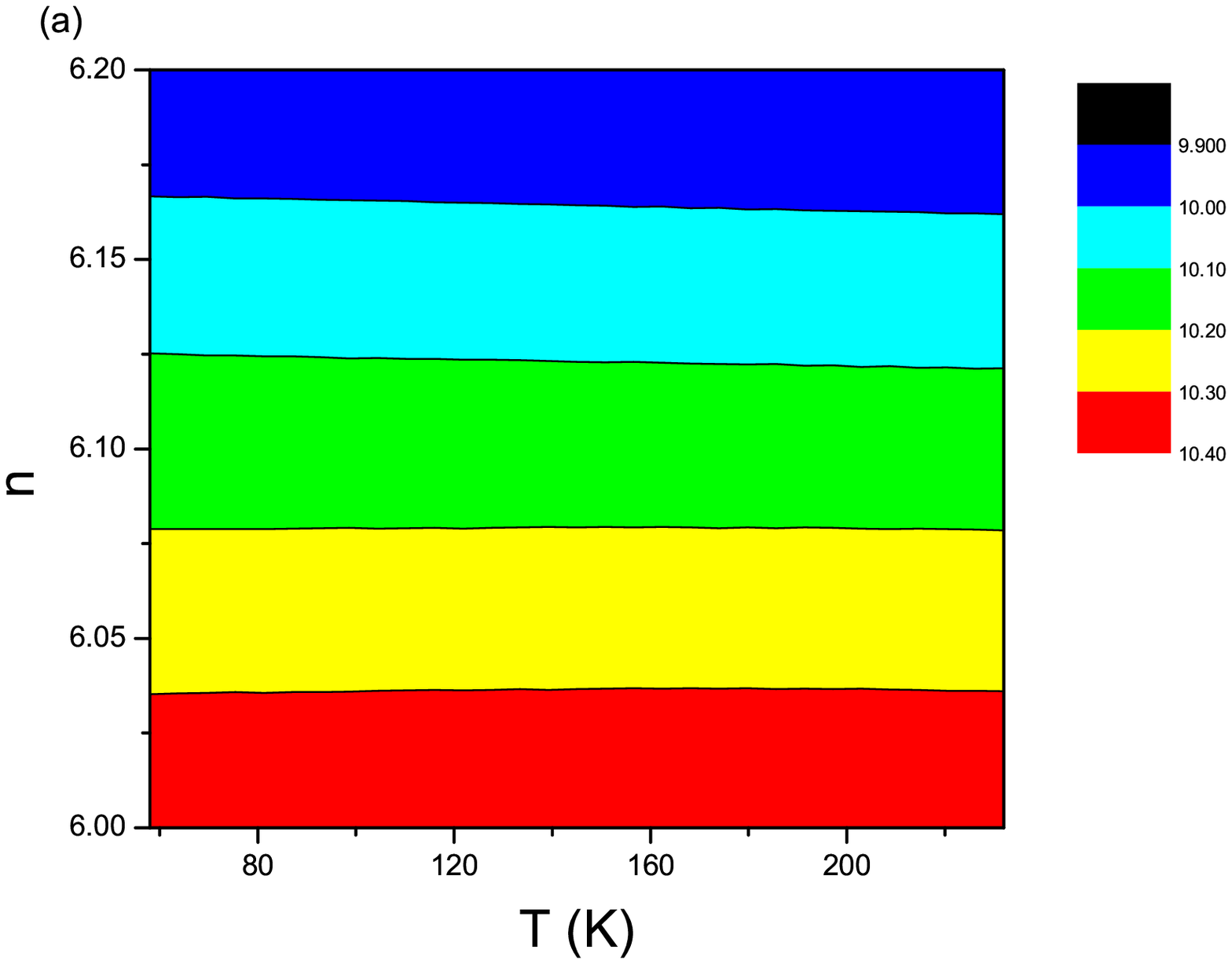}
\includegraphics[width=7cm,angle=0]{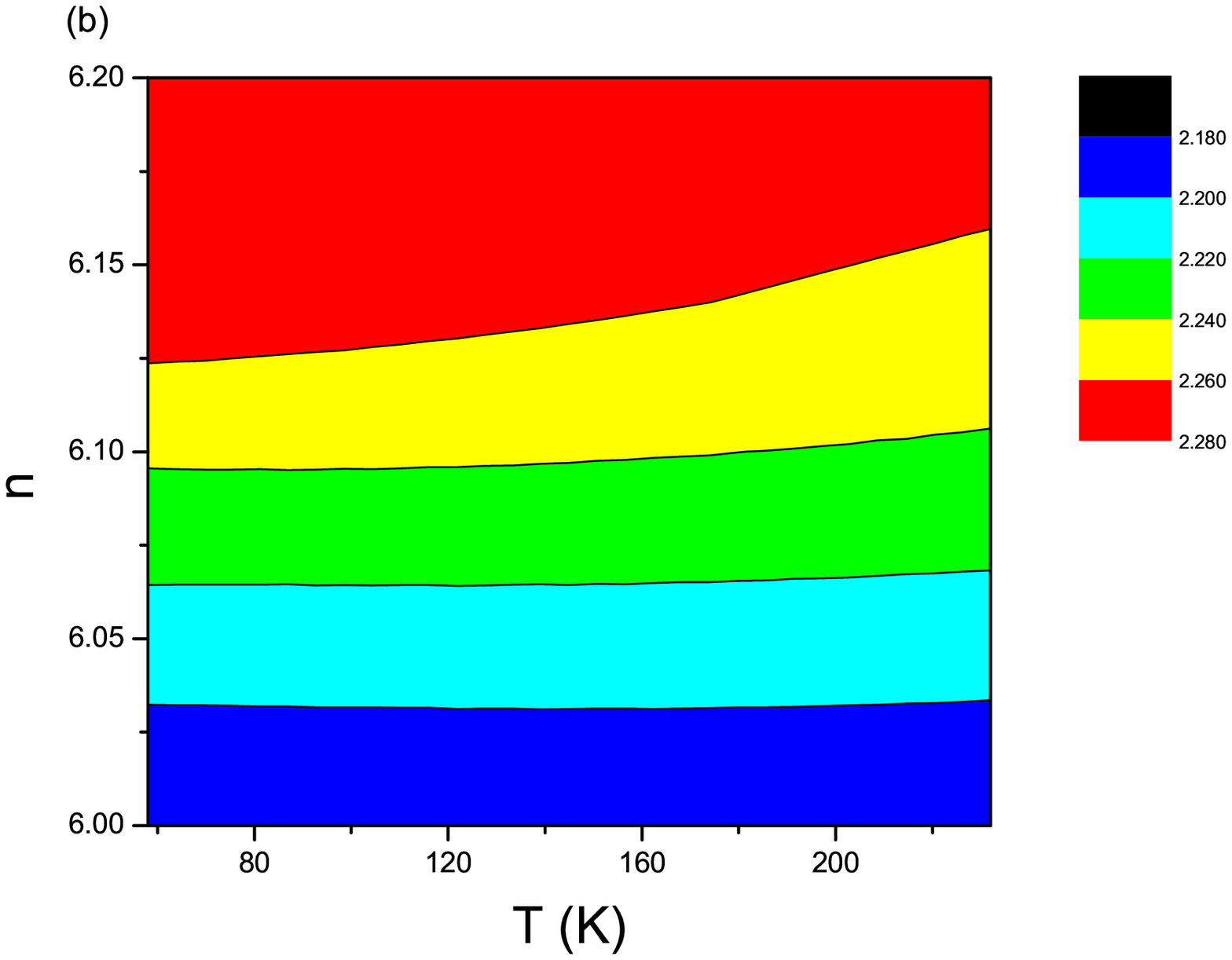}
\caption{The temperature and doping dependence of the RPA-corrected
in-plane orbital magnetic susceptibility(a) and the ratio between
the in-plane and out-of-plane orbital magnetic susceptibility(b) for
$6.0\leq n\leq6.2$.} \label{sfig2}
\end{figure}

\subsection{The temperature and doping dependence of the anisotropy ratio}
Unlike the spin magnetic response, the orbital magnetic response is
contributed by both intra-band and inter-band process. As a result,
the orbital magnetic response is much less sensitive to the
variation of temperature and doping concentration of the system. In
Fig.\ref{sfig2}, we present the temperature and doping dependence of
the RPA-corrected in-plane orbital magnetic susceptibility and the
ratio between the in-plane and the out-of-plane orbital magnetic
susceptibility.

From the figure it is clear that both quantities have only small
temperature and doping dependence. More specifically, the relative
change of the in-plane orbital magnetic susceptibility for $6.0\leq
n\leq6.2$ is only about 5 percent. The change in the anisotropy
ratio is less than 0.1.


\begin{thebibliography}{99}
\bibitem{group}Here $R_{x}(\pi)$ and $R_{y}(\pi)$ are the $\pi$ rotations along the $x$ and $y$ axis, $\sigma_{X}$ and $\sigma_{Y}$ are the mirror
planes normal to the $X$ and $Y$ axis. $R_{x}(\pi)$, $\sigma_{X}$
and $\sigma_{Y}$ form a set of generators of the symmetry group
$D_{2d}$ of the tetragonal phase and $R_{x}(\pi)$ and $R_{y}(\pi)$
form a set of generators of the symmetry group $D_{2}$ of the
orthogonal phase.
\end{thebibliography}
\end{document}